\documentclass[aps,prb,preprint, superscriptaddress,showpacs]{revtex4}  
\usepackage{graphicx}  
\usepackage{dcolumn}   
\usepackage{bm}        
\usepackage{amssymb}   
\usepackage[dvipsnames,usenames]{color}
\usepackage{color}
\usepackage{pdfpages}

\begin{document}


\newcommand{\etal}{{\sl et al.}}
\newcommand{\ie}{{\sl i.e.}}
\newcommand{\sto}{SrTiO$_3$} 
\newcommand{\lto}{LaTiO$_3$}
\newcommand{\lao}{LaAlO$_3$}
\newcommand{\lno}{LaNiO$_3$}
\newcommand{\ngo}{NdGaO$_3$}
\newcommand{\gfo}{GdFeO$_3$}
\newcommand{\tith}{Ti$^{3+}$}
\newcommand{\tifo}{Ti$^{4+}$}
\newcommand{\otw}{O$^{2-}$}
\newcommand{\alo}{AlO$ _2 $}
\newcommand{\tio}{TiO$ _2 $}
\newcommand{\eg}{$e_{g}$}
\newcommand{\tg}{$t_{2g}$}
\newcommand{\dzt}{$d_{z^2}$}
\newcommand{\dxtyt}{$d_{x^2-y^2}$}
\newcommand{\dxy}{$d_{xy}$}
\newcommand{\dxz}{$d_{xz}$}
\newcommand{\dyz}{$d_{yz}$}
\newcommand{\egp}{$e'_{g}$}
\newcommand{\ag}{$a_{1g}$}
\newcommand{\mub}{$\mu_{\rm B}$}
\newcommand{\ef}{$E_{\rm F}$}
\newcommand{\alao}{$a_{\rm LAO}$}
\newcommand{\ango}{$a_{\rm NGO}$}
\newcommand{\asto}{$a_{\rm STO}$}
\newcommand{\nst}{$N_{\rm STO}$}
\newcommand{\lamstn}{(LAO)$_M$/(STO)$_N$}
\newcommand{\ngmstn}{(NGO)$_M$/(STO)$_N$}

\parindent 0pt 

\title{Control of orbital reconstruction in (LaAlO$_3$)$_M$/(SrTiO$_3$)$_N$(001) quantum wells by strain and confinement}

\author{David Doennig}
\affiliation{Forschungs-Neutronenquelle Heinz Maier-Leibnitz (FRM II), Technische Universit\"at M\"unchen, Lichtenbergstra\ss{}e 1, 85748 Garching, Germany}
\author{Rossitza Pentcheva}
\affiliation{Department of Physics and Center for Nanointegration Duisburg- Essen (CENIDE),
University of Duisburg-Essen, Lotharstr. 1, 47057 Duisburg, Germany}
\affiliation{Forschungs-Neutronenquelle Heinz Maier-Leibnitz (FRM II), Technische Universit\"at M\"unchen, Lichtenbergstra\ss{}e 1, 85748 Garching, Germany}
\date{\today}

\begin{abstract}
The diverse functionality emerging at oxide interfaces calls for a fundamental understanding of the mechanisms and control parameters of electronic reconstructions.  Here, we explore the evolution of electronic phases in (LaAlO$_3$)$_M$/(SrTiO$_3$)$_N$(001) superlattices as a function of strain and confinement of the SrTiO$_3$ quantum well. Density functional theory calculations including a Hubbard $U$ term reveal a charge ordered \tith\ and \tifo\ state for $N=2$ with an unanticipated orbital reconstruction, displaying alternating \dxz\ and  \dyz\ character at the \tith\ sites, unlike the previously reported \dxy\ state, obtained only for reduced $c$-parameter at \asto. At \alao\ $c$-compression leads to a Dimer-Mott insulator with alternating \dxz, \dyz\ sites and an almost zero band gap. Beyond a critical thickness of $N=3$ (\asto) and $N=4$ (\alao) an insulator-to-metal transition takes place, where the extra $e/2$ electron at the interface  is redistributed throughout the STO slab with a \dxy\ interface orbital occupation and a mixed \dxz\ + \dyz\ occupation in the inner layers. Chemical variation of the \sto\ counterpart (\lao\ vs. NdGaO$_3$) proves that the significant octahedral tilts and distortions in the \sto\ quantum well are induced primarily by the electrostatic doping at the polar interface and not by variation of the \sto\ counterpart. 

\bigskip\noindent
$^{*}$Correspondence to: Rossitza.Pentcheva@uni-due.de
\end{abstract}

\maketitle

Oxide interfaces show surprisingly rich electronic behavior that is not present in the bulk compounds. Among the different systems, the (001)-oriented \lao/\sto\ interface has been most prominent, displaying two-dimensional conductivity, superconductivity and magnetism, as well as confinement-induced metal-to-insulator transitions~\cite{ohtomo,reyren,brinkman,thiel,hwang2012}. The majority of experimental studies have explored thin \lao\ (LAO)-films on \sto(001). In contrast, superlattices containing two $n$-type interfaces with a LaO and TiO$_2$ stacking have been in the focus of theoretical studies~\cite{pentcheva2006,pentcheva2008,zhong2008,popovich,tsymbal}. The latter predict that the excess charge of 0.5$e$ at the polar interface results either in a charge ordered phase with  \dxy\ orbital polarization or a quasi 2DEG where the excess charge of 0.5$e$ is delocalized throughout the \sto\ (STO) quantum well. While the experimental realization of this type of superlattice remains a challenge as it requires precise control of termination at both interfaces, there is renewed interest in such systems in view of confinement effects in electrostatically doped oxide quantum wells. To this end, recent experimental studies on related GdTiO$_3$/\sto\ superlattices  by Moetakef \emph{et al.} \cite{moetakef2012} reported the formation a high density 2DEG  where the carrier concentration can be modulated by the thickness of the STO quantum well. Moreover, a metal-to-insulator transition and ferromagnetism is induced for STO thickness below two unit cells (containing two SrO layers)~\cite{moetakefPRX}. In contrast, the analogous SmTiO$_3$/\sto\ superlattice (SL) remains metallic down to a single SrO layer~\cite{zhang2014}. This distinct behavior of the two systems has been related to differences in octahedral tilts and distortions, induced by GdTiO$_3$ or SmTiO$_3$.

These findings have motivated us to revisit the LAO/STO superlattices and explore systematically the types of electronic and orbital reconstructions that can be realized as a function of STO confinement. We note that LAO/STO SLs have the advantage that the transition metal ion is only confined to the STO part, therefore emerging effects, including magnetism can be unambiguously associated with the STO quantum well (QW). We relate the electronic behavior to the structural changes by fully considering octahedral tilts and distortions as well as the role of the vertical lattice parameter. We note that previous work~\cite{pentcheva2006, pentcheva2008} has addressed predominantly tetragonally distorted superlattices, where due to symmetry restrictions, only a structural optimization of the $z$ position of the ions was performed. GdFeO$_3$-type distortions have been considered by Zhong and Kelly \cite{zhong2008} only for a \lamstn(001) superlattices with $N=3$. Here, we have investigated the behavior as a function of $N$. In particular, a  critical thickness for a metal-to-insulator transition in this system is determined.

Strain is a powerful tool to tune the electronic behavior and orbital polarizations {\cite{schlom,rondinelli}, but so far only a few studies have addressed this effect in the LAO/STO system. Bark \emph{et al.} \cite{bark2011} studied its influence on the properties of \lao\ films on \sto(001) by growing the system on substrates with different lattice parameters. Transport measurements show that under compressive strain the 2DEG is preserved, but with reduced carrier concentration compared to the system grown on \sto(001), whereas for larger tensile strain the 2DEG is quenched. 
Furthermore, recent experiments have shown variations of carrier density and mobility of the 2DEG when exchanging \lao\ by other polar $RB$O$_3$ oxides ($R$=La, Pr, or Nd, $B$=Al or Ga) \cite{ariandoprb86}.

\begin{figure*}[h!tbp]
\includegraphics[scale=0.65]{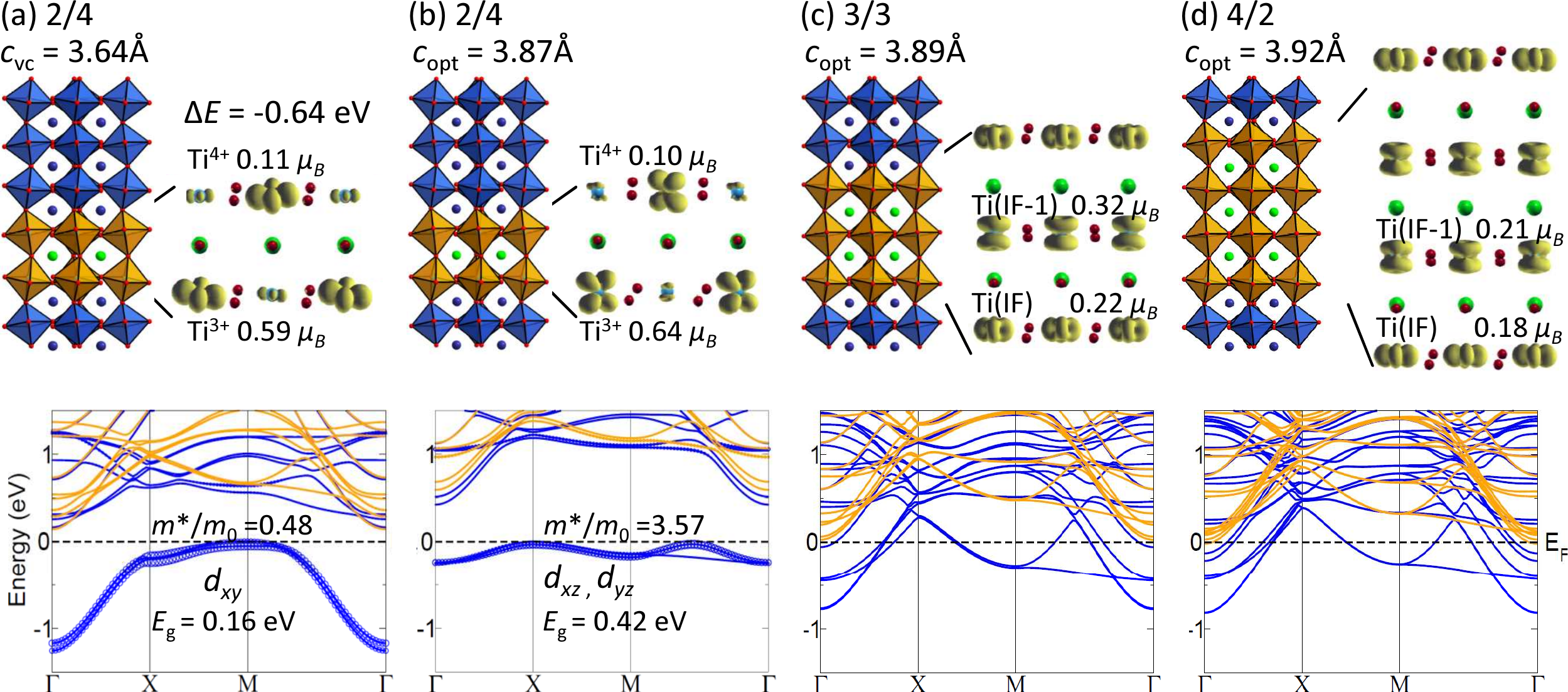}
\caption{
Side view of the relaxed structure, electron density distribution, integrated over occupied Ti 3$d$ bands between \ef-1.5 eV  and \ef\ and band structure in \lamstn\ (001) SL at \asto. Majority and minority bands are plotted in blue and orange, respectively. Character plotting highlights the occupied $d$ band at the \tith\ sites. (a) $N=2$ is a charge-ordered FM insulator with \dxy\ orbital occupation at the \tith\ sites for the $c$-parameter fixed to conserve the volume and (b) with \dxz\ and \dyz\ orbital occupation at \tith\ at each interface layer for the optimized $c$-parameter. $\Delta E$ is the energy difference (in eV per Ti site) between the cases shown in (b) and (a);(c)-(d) For $N=3, 4$ STO layers, the charge-ordered state is suppressed and the system switches from insulating to conducting behavior with a \dxy\ orbital at the Ti interface sites and \dxz\ + \dyz\ orbital occupation in the central STO layers.}
\label{fig:Fig01}
\end{figure*}

The goal of this theoretical study is to explore systematically the influence of strain, confinement and chemical variation on the electronic and orbital reconstruction in (001)-oriented \lamstn\ superlattices, where $N$ is varied between 2-4.  Octahedral tilts and distortions, considered to be an important factor in tuning the properties of the 2DEG at these oxide interfaces, were fully taken into account. Additionally, we also performed optimization of the out-of-plane lattice parameter $c$. The effect of chemical control on the interface properties is addressed by replacing the polar \lao\ with the polar \ngo (NGO).

\begin{figure*}[h!tbp]
\includegraphics[scale=0.65]{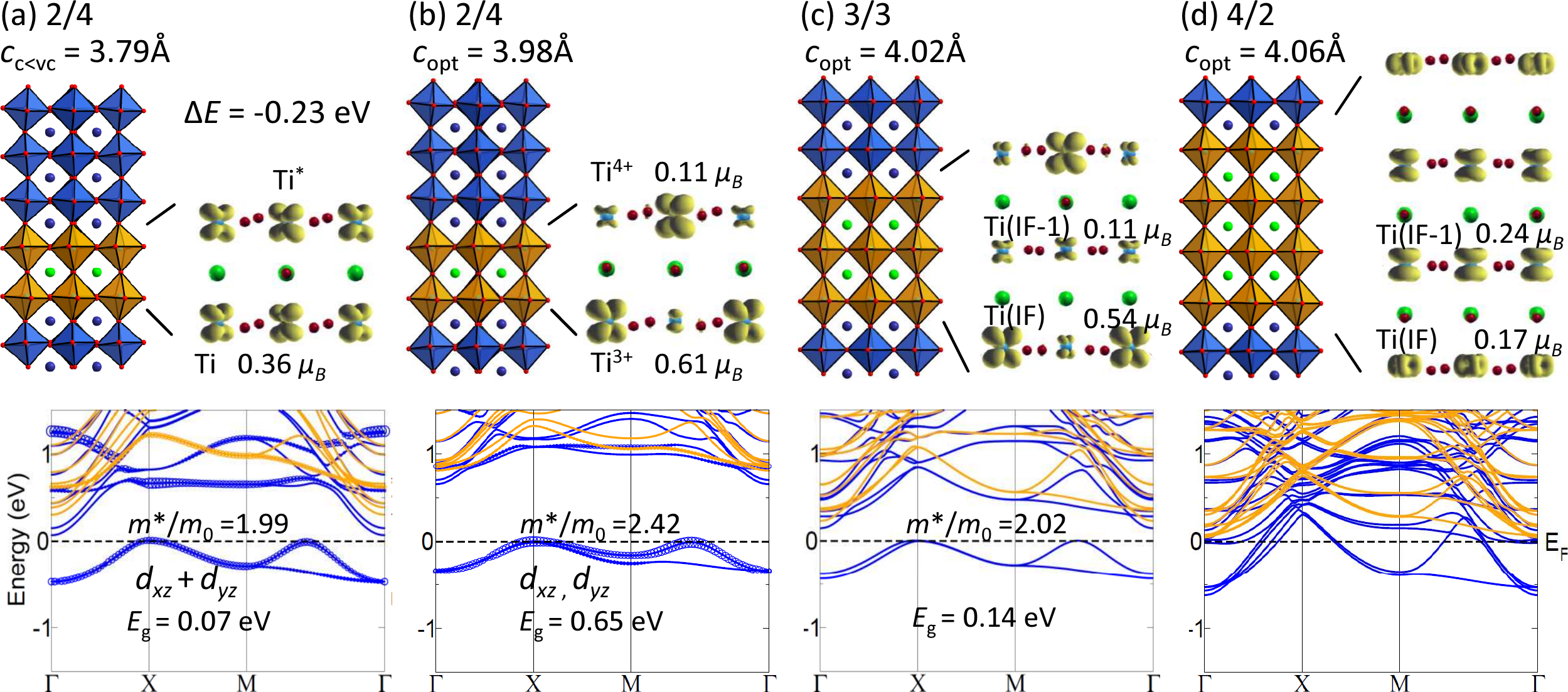}
\caption{
Side view of the relaxed structure, electron density distribution, integrated over occupied Ti 3$d$ bands between \ef-1.5 eV  and \ef, and band structure of  \lamstn\ (001) SL at \alao, $N=2-4$. Majority and minority bands are plotted in blue and orange, respectively. Character plotting highlights the occupied $d$ band at the \tith\ sites. (a) For $N=2$ a Dimer Mott insulator is obtained for the strongly compressed $c$ parameter with alternating \dxz, \dyz\ orbital occupation at the Ti$^{3.5+}$ sites; (b) for the optimized $c$ a charge ordered state with \dxz\ and  \dyz\ orbital occupation at the \tith\ sites at each interface layer emerges; $\Delta E$ is the energy difference (in eV per Ti site) between the cases shown in (b) and (a);
 (c) $N=3$ is similar to (b) with \tifo\ in the central layer; (d) an insulator to metal transition takes place for $N=4$:   with a \dxy\ orbital polarization at the Ti interface sites and \dxz\ + \dyz\ orbital occupation in the inner STO layers.}
\label{fig:Fig02}
\end{figure*}

\newpage
{\Large {\bf Results}}\\\\
{\large {\bf Electronic and orbital reconstruction \lamstn(001) SLs: influence of strain}}
\label{subsec:reconstruction}

{\it Results for \asto, corresponding to an underlying STO(001) substrate.}

Figure \ref{fig:Fig01} presents results for $n$-type (001)-oriented \lamstn\ superlattices with varying thickness of the STO quantum well, $N=2-4$. 
Using the $c$ parameter that corresponds to volume conservation of the corresponding bulk constituents, the \sto\ bilayer ($N=2$ case) under tensle strain (\asto) [Fig. \ref{fig:Fig01}(a)] is a charge-ordered ferromagnetic insulator with alternating \tith\ (0.59\,$\mu_B$) and \tifo\ (0.11\,$\mu_B$) sites with a \dxy\ orbital occupation at the \tith\ sites, consistent with previous work \cite{pentcheva2006,pentcheva2008,zhong2008}. Surprisingly, for the optimized $c$ parameter of 3.87\,\AA\ (0.23\,\AA\ larger) the charge-ordered pattern is preserved, however the orbital order at the \tith\ sites changes to \dxz\ for  \tith\ at the top interface  and \dyz\ for \tith\ at the bottom interface [Fig. \ref{fig:Fig01}(b)]. This is associated with a much narrower occupied Ti 3$d$ band and an enhanced effective mass of 3.57\,$m_e$, indicating an increase of electron correlation effects. Simultaneously the size of the band gap is enlarged from 0.16\,eV to 0.42\,eV. The magnetic moment of the \tith\ sites increases slightly from 0.59\,$\mu_B$ to 0.64\,$\mu_B$. 

The mechanism of electronic reconstruction is distinct for  the thicker STO quantum wells, $N=3$ [Fig. \ref{fig:Fig01}(c)] and  $N=4$ [Fig. \ref{fig:Fig01}(d)]. Here,  the extra $e$/2 charge from each interface is redistributed  throughout the STO slab, resulting in an insulator-to-metal transition with a critical thickness $N_c=3$. In both cases the charge-ordered state is suppressed. Simultaneously, the orbital polarization is altered to predominantly \dxy\ orbital character in the interface Ti-layer and mixed \dxz\ + \dyz\ in the central Ti layers, similar to  previous findings in \lamstn\ superlattices \cite{pentcheva2008,popovich,tsymbal,Salluzzo}.

{\it Results for \alao, corresponding to an underlying LAO(001) substrate.}

Figure \ref{fig:Fig02} shows the results for (001)-oriented \lamstn\ heterostructures with STO thickness ranging from $N=2-4$. 
For the \sto\ bilayer ($N=2$) the ground state under compressive strain (\alao) and optimized $c$ parameter [see Fig. \ref{fig:Fig02}(b)] is analogous to the one for \asto\ [cf. Fig. \ref{fig:Fig01}(b)]:  a charge-ordered insulating phase with \tith\ and \tifo\ in a checkerboard arrangement and a  \dxz, \dyz\ orbital polarization at the \tith\ sites at each interface, albeit with slightly smaller magnetic moment (0.61\,$\mu_B$) at the \tith\ sites. This is accompanied by an increase of the band gap from 0.42\,eV (\asto) to 0.65\,eV (\alao). 

An interesting electronic state  occurs for $N=2$ upon compression of the  $c$ parameter of the superlattice to the bulk LAO value [see Fig. \ref{fig:Fig02}(a)]: The charge-ordered state is quenched, leaving 0.5$e$ at each Ti site (0.33\,$\mu_B$) and alternating \dxz, \dyz\ orbital polarization on neighboring Ti sites in a checkerboard arrangement. Characteristic for this state is a decreased but nonzero band gap (0.07\,eV). This so called Dimer Mott insulator (DMI), initially predicted within the framework of one-dimensional Hubbard ladders~\cite{dagotto1996}, was recently predicted for GTO/STO QWs~for a single SrO layer in a GdTiO$_3$ matrix\cite{chen2013}. 
In fact the ground state for the optimized $c$-value can be regarded as emerging from the DMI phase by transferring 0.5$e$ between each pair of Ti sites, leading to a \tith\ and \tifo\ charge order with some residual \dxz\ orbital occupation at the \tifo\ sites and \dyz\ at the \tith\ sites at one interface and the opposite orbital polarization in the  other interface layer.          

To explore how the combination of exchange correlation functional (LDA vs GGA) and the Coulomb repulsion term influence the structural properties and electronic ground state, we have performed additional calculations for (LAO)$_4$/(STO)$_2$(001) at \alao, using the LDA+$U$ functional. Setting the lateral lattice parameter to the experimental bulk LAO value of 3.79\,\AA, the optimized $c$ lattice parameter (3.83\,\AA) is $\sim 4\%$ smaller than the GGA+$U$ value (3.98\,\AA). The electronic ground state in Fig. \ref{fig:Fig02}(b) with \tith\ and \tifo\ in a checkerboard arrangement and a  \dxz, \dyz\ orbital polarization at the \tith\ sites emerges within LDA+$U$ beyond $U=5.5$ eV. Thus we can conclude that the electronic ground state is largely unaffected by the reduction of the $c$-lattice parameter within LDA+$U$ and is stabilized for a slightly higher value of $U$ than in the GGA+$U$ calculation.

Unlike the corresponding case at \asto, the electronic and orbital reconstruction  persists for  $N=3$ [see Fig. \ref{fig:Fig02}(c)] i.e. a charge-ordered state with alternating \dxz, \dyz\ at the \tith\ sites emerges at both interface layers, whereas the central layer becomes \tifo. Compared to the case with $N=2$ with optimized $c$, the band width is increased and the band gap is reduced to 0.14\,eV. The magnetic moment at the \tith\ sites is reduced to (0.54\,$\mu_B$), indicating an enhanced delocalization.    

An insulator-to-metal transition takes place at a STO thickness $N=4$ [see Fig. \ref{fig:Fig02}(d)]. The suppression of the charge ordered state leads to a closing of the band gap. The extra $e/2$ charge is delocalized throughout the whole STO quantum well with a preferential \dxy\ orbital occupation at the interface layer and a \dxz\ + \dyz\ occupation in the central layers.

\bigskip
{\large {\bf Structural relaxations}}\bigskip
\label{subsec:relaxation}

To investigate the relationship between electronic properties and structural distortions we have performed a quantitative analysis of the fully relaxed \lamstn\ (001) superlattices. The B-O-B bond angles in the  perovskite structure ABO$_3$ are closely related to octahedral tilts and determine orbital overlaps, and are therefore considered to be the control parameter for metal-to-insulator transitions in bulk transition metal oxides~\cite{goodenough,medarde}. Understanding the origin of modifications of octahedral tilts and connectivity at interfaces in oxide superlattices is an emerging topic of considerable interest~\cite{rondinelli,spaldin}. 

\begin{figure}[h!tbp]
\includegraphics[scale=0.71]{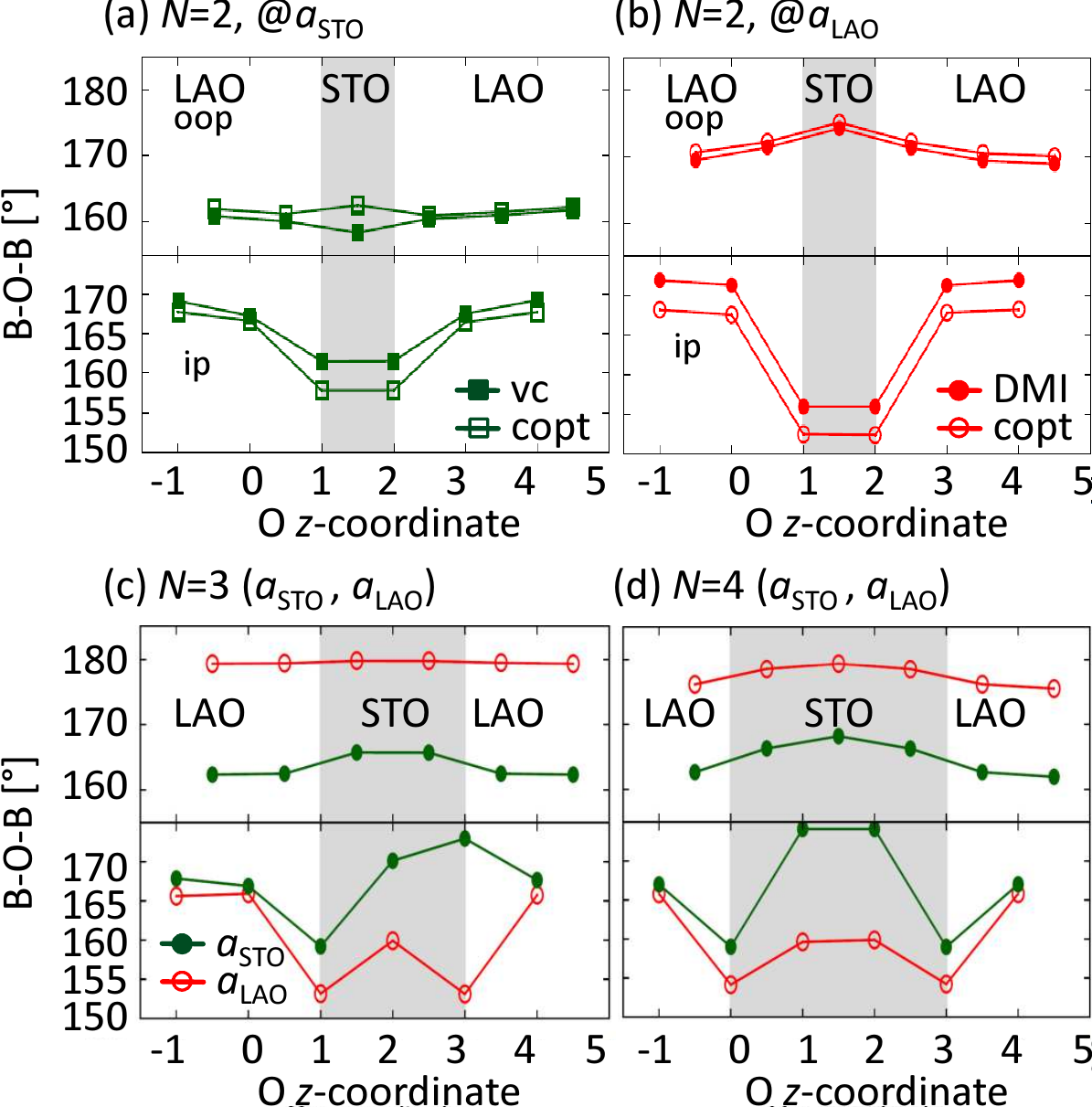}
\caption{
Quantitative analysis of B-O-B bond angle in \lamstn\ (001) superlattices for different STO thicknesses $N$ (shaded region). (a) The bilayer $N=2$ under tensile strain (\asto). (b) STO bilayer for compressive strain (\alao). (c)-(d) $N=3,4$ STO layers.}
\label{fig:Fig03}
\end{figure}

In Fig. \ref{fig:Fig03} we present the B-O-B bond angles of the superlattices as a function of the in-plane lattice constant \asto\ and \alao. The corresponding  results for \ango\ are displayed in Fig. \ref{fig:Fig04}. In the \lamstn\ (001) superlattices we identify strong reduction  of the B-O-B angles from the bulk values. We remind that bulk \sto\ has the ideal cubic perovskite structure ($Pm\overline{3}m$) with B-O-B bond angles of 180$^\circ$. Below 105\,K it undergoes an antiferrodistortive phase transition to a tetragonal ($I4/mcm$) structure~\cite{burkhard1979}, whereas \lao\ attains a rhombohedral structure ($R\overline{3}c$) at all temperatures with B-O-B bond angles of 171.4$^\circ$. The deviations of the B-O-B bond angles from the ideal 180$^\circ$ are most prominent in the STO quantum well. Moreover, there are pronounced differences in the behavior of in-plane and out-of-plane angles and their strain dependence: For all studied thicknesses of the STO quantum well ($N=2-4$), the in-plane B-O-B bond angles show the strongest deviation from 180$^\circ$ in particular in the STO part, while in the LAO part the deviations of the bond angles are smaller and converge towards the bulk LAO value away from the interface. As a function of strain, the ones in the STO part ($N=2$) increase with the lateral lattice constant, i.e. $\sim 152^\circ$ at \alao, followed by $\sim 155^\circ$ at \ango\  and finally $\sim 158^\circ$ at \asto.  The opposite trend is observed for the out-of-plane BOB bond angles. Here the strongest deviation from 180$^\circ$ is for \asto\ ($\sim 162^\circ$), then \ango\ ($\sim 168^\circ$) and lowest for \alao\ ($\sim 175^\circ$). Moreover, the variation in out-of-plane bond angles between the LAO and STO part of the SL is much weaker than the in-plane bond angles. 

\begin{figure}[h!tbp]
\includegraphics[scale=0.67]{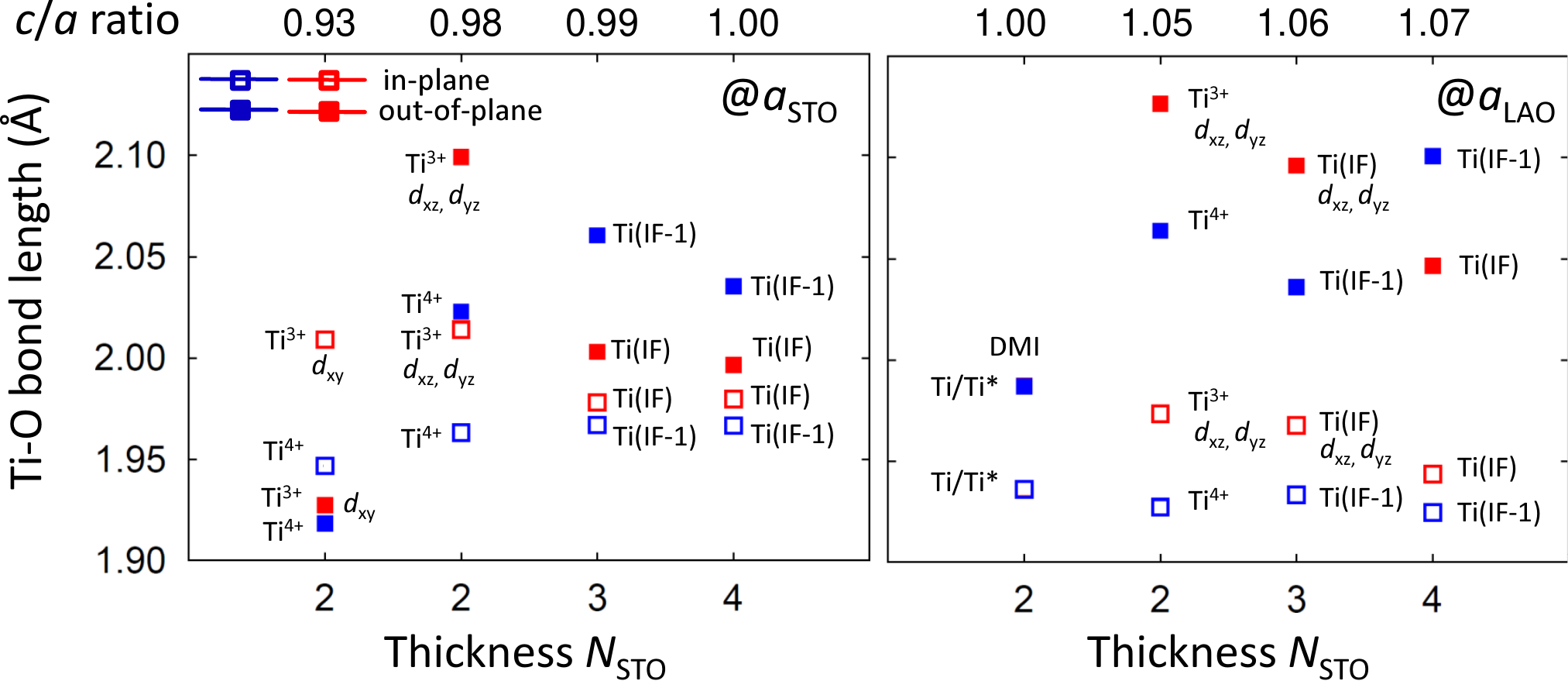}
\caption{
In-plane (hollow squares) and out-of-plane (solid squares) Ti-O bond lengths for the distinct Ti sites in \lamstn\ (001) superlattices at \asto (left panel) and \alao (right panel). The corresponding $c/a$ ratio is given at the top of the graphs.}
\label{fig:Fig06}
\end{figure}

In order to relate the structural distortions to the resulting orbital reconstruction, in Fig. \ref{fig:Fig06} we have also analyzed the corresponding Ti-O bond lengths. $N=2$ under tensile strain with $c$-parameter fixed to conserve the volume comprises the only case where the in-plane bond distance is larger than the out-of-plane Ti-O length. This is consistent with the \dxy\ orbital order at \tith. For all other cases, the out-of-plane Ti-O bond distance is larger than the in-plane value. These differences are particularly pronounced at the \tith\ sites, e.g. 2.10 vs. 2.01\,\AA\ at \asto\ and 2.13 vs. 1.97  at \alao\ for $N=2$, respectively. A similar trend but with smaller absolute values is observed  at the \tifo\ sites (2.02 vs. 1.96\,\AA\ and 2.06 vs. 1.93\,\AA, respectively). The enhanced out-of-plane to in-plane distance ratio favors the alternating \dxz, \dyz\ orbital polarization. As this trend is stronger for \alao, it is responsible for the stabilization of this orbital order and consequently the  insulating state up to $N=3$, while at \asto\ the system is already metallic for $N=3$. As discussed above, the insulator-to-metal transition is connected with a change in orbital polarization to \dxy\  in the interface layer and mixed \dxz+\dyz\ in the inner layer. Indeed the difference in out-of-plane to in-plane distance is reduced (\alao: 2.06 vs. 1.93 \AA) and almost quenched at Ti(IF) (\asto: 2.00 vs 1.98 \AA) for $N=4$. Conversely, the inner (IF-1) layers show a larger difference (\alao: 2.10 vs. 1.92 \AA; \asto: 2.03 vs. 1.97 \AA). Thus changes in the in-plane and out-of-plane bond lengths clearly reflect the trends in orbital polarization as a function of strain and thickness $N$ of the STO quantum well.

\bigskip
{\large {\bf Chemical control: \lamstn(001) vs. \ngmstn(001)}}\bigskip
\label{sec:chemical}

To explore how chemical variation of the STO counterpart influences the electronic reconstruction and relaxations in the STO QW, we performed calculations on $n$-type \ngmstn\ (001) superlattices with the in-plane lattice parameter fixed to \ango, which formally has the same polar discontinuity across the interface. For comparison, Figs. \ref{fig:Fig04}, \ref{fig:Fig05a} and \ref{fig:Fig05b} contain also the structural and electronic properties  \lamstn(001) superlattices at \ango.

\begin{figure}[h!tbp]
\includegraphics[scale=0.71]{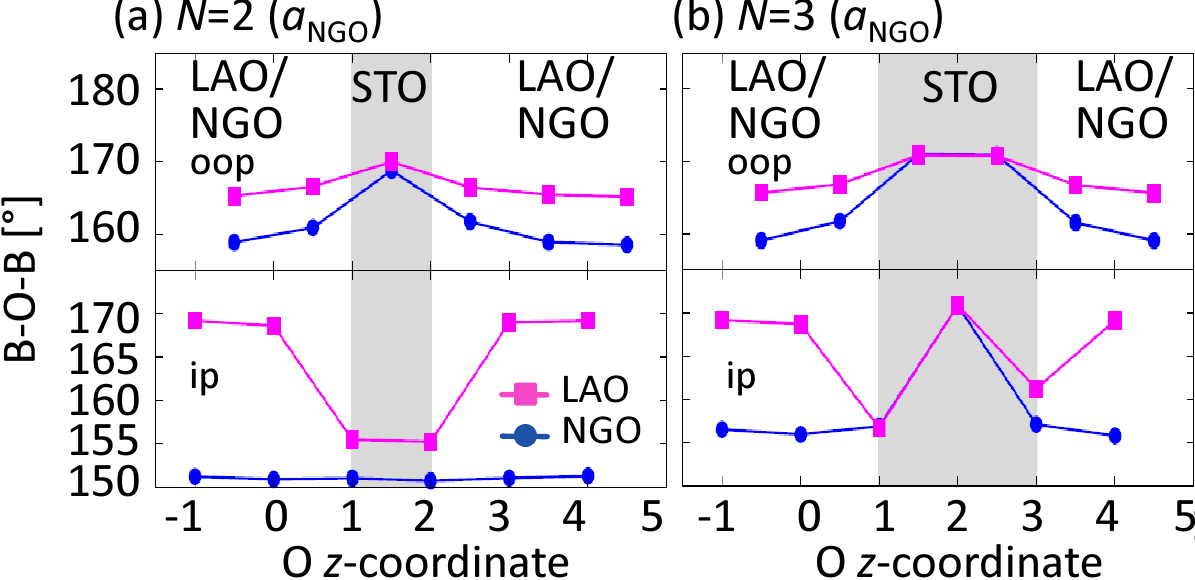}
\caption{
Quantitative analysis of B-O-B bond angle in (001)-oriented \lamstn\ and \ngmstn\ superlattices with the in-plane lattice parameter fixed at \ango\ for different STO thicknesses $N$ (shaded region). (a) $N=2$ STO layers. (b) $N=3$ STO layers.}
\label{fig:Fig04}
\end{figure}

Fig. \ref{fig:Fig04} compares the evolution of in-plane and out-of-plane B-O-B bond angles across the interface, for the \ngo/\sto\ and \lao/\sto\ superlattices with $N=2, 3$. For $N=2$ (see Fig. \ref{fig:Fig04}(a)), the in-plane bond angle in the \ngo/\sto\ heterostructure remains nearly constant ($\sim 151^\circ$) throughout the superlattice. This value is slightly smaller but close to the bulk value (153.2$^\circ$) of orthorhombic NGO ($Pbnm$). For the \lao/\sto\ superlattice the bond angle in the STO quantum well is strongly reduced from the bulk STO value to $\sim 155^\circ$. Moving away from the interface into the LAO part, the B-O-B bond angles quickly approach the LAO bulk value of 171.4$^\circ$. In contrast, for both superlattices, the out-of-plane bond angle deviations 180$^\circ$ in the STO part are less pronounced and the values almost coincide for the  \ngo/\sto\  and the  \lao/\sto\ superlattices in the center of the STO part, whereas they converge towards the NGO and LAO bulk values outside the STO QW.  

\begin{figure}[h!tbp]
\includegraphics[scale=0.67]{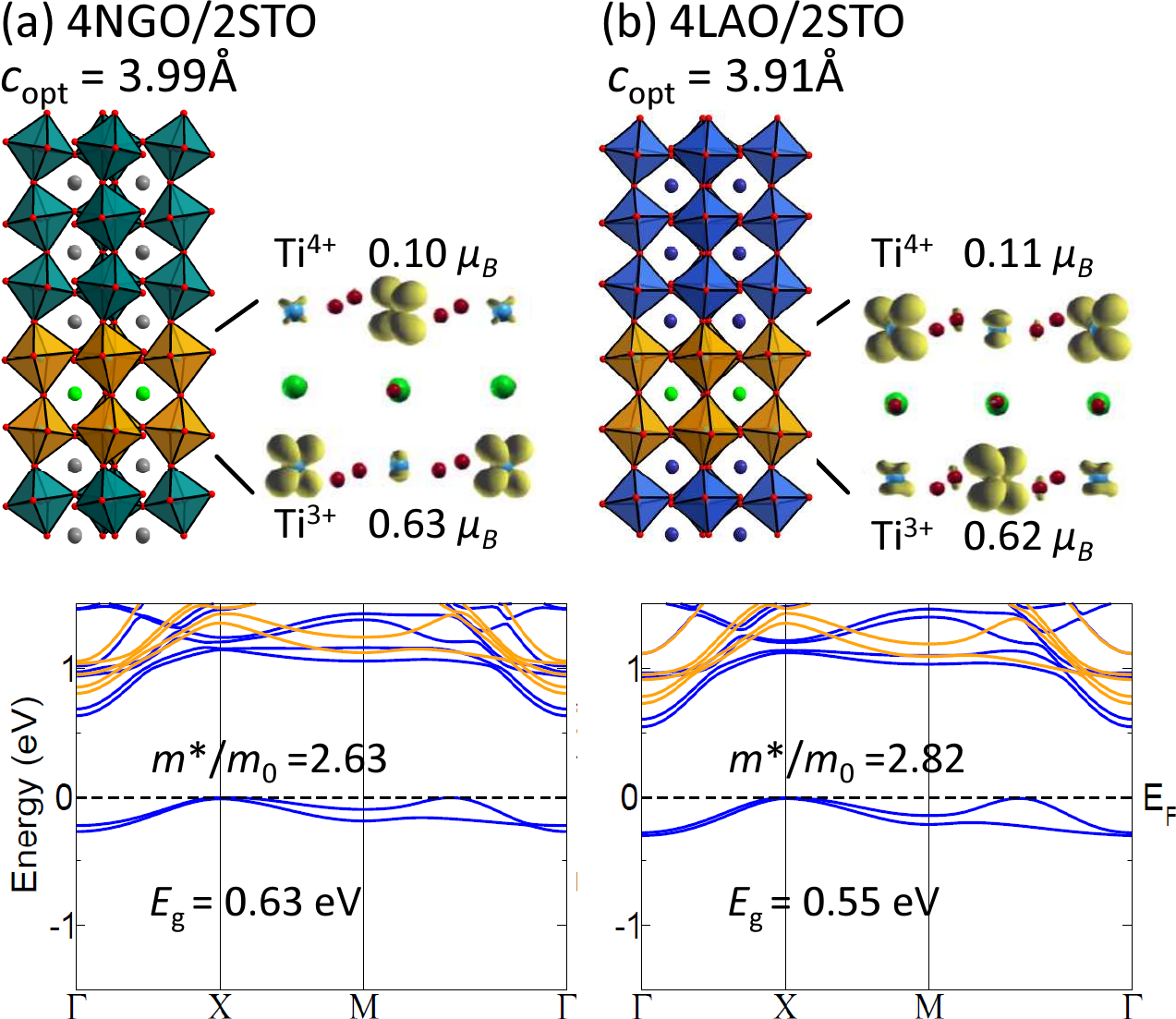}
\caption{
Side view of the relaxed structure, electron density distribution, integrated over occupied Ti 3$d$ bands between \ef-1.5 eV  and \ef\ and band structure in (001)-oriented \ngmstn\ and \lamstn\ superlattices at \ango\ with $N=2$. Majority and minority bands are plotted in blue and orange, respectively.}
\label{fig:Fig05a}
\end{figure}

\begin{figure}[h!tbp]
\includegraphics[scale=0.67]{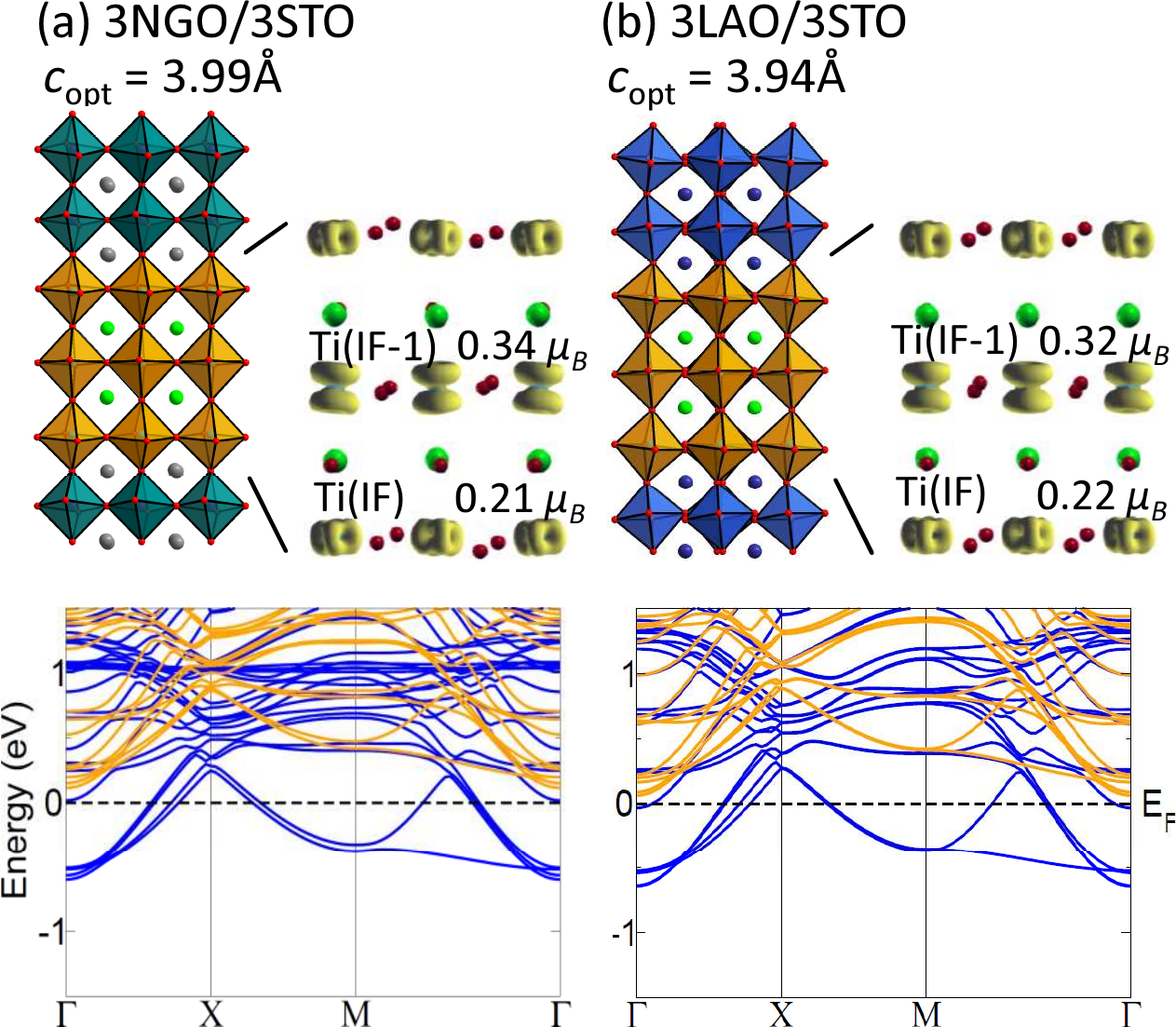}
\caption{
Side view of the relaxed structure, electron density distribution, integrated over occupied Ti 3$d$ bands and band structure in (001)-oriented \ngmstn\ and \lamstn\ superlattices at \ango\ with $N=3$. Majority and minority bands are plotted in blue and orange, respectively.}
\label{fig:Fig05b}
\end{figure}

Increasing the STO thickness to $N=3$ (see Fig. \ref{fig:Fig04}(b)), the in-plane bond angle deviations are significantly reduced in the central STO part ($\sim 170^\circ$), but are still quite prominent at the interface ($\sim 157^\circ$). The deviation of the out-of-plane bond angles is similar as for the superlattices with $N=2$.   

The similarity in structural relaxation of the  \ngo/\sto\  and the  \lao/\sto\ superlattices goes hand in hand with common electronic properties: 
For $N=2$, the ground state for both systems is an insulating charge-ordered phase with \tith\ and \tifo\ sites (see Fig. \ref{fig:Fig05a}). Increasing the STO thickness to $N=3$, the system undergoes an insulator-to-metal transition (see Fig. \ref{fig:Fig05b}). The charge-ordered state is suppressed and the excess charge is delocalized throughout the STO slab, similar to the corresponding case at \asto\ (see Fig. \ref{fig:Fig01}(c)).

\bigskip
{\Large {\bf Discussion}}\bigskip

A central question is the origin of the unanticipated orbital reconstruction and how it relates to the STO QW thickness and structural distortions. Our results suggest that the electronic ground state and orbital polarization are intimately related to the level of electrostatic doping of the system. By construction the polar discontinuity at the interface introduces $0.5e$ per lateral unit cell at each interface, corresponding to a carrier density of $\sim 7 \times10^{14}$ cm$^{-2}$. $N=2$ comprises an extreme case with doping of $0.5 e$ per Ti site. With increasing $N_{\rm STO}$ the excess charge is redistributed throughout the STO QW, resulting in a reduction of the carrier density and the effective filling of the Ti $3d$ band.   We consider this reduction of carrier density with increasing $N_{\rm STO}$ as the driving force for the switching of orbital polarization from \dxz, \dyz\ at the \tith\ sites for $N=2$ to \dxy\ in the interface layer and mixed \dxz, \dyz\ in the inner layers beyond $N=3$ as well as for the concomitant insulator-to-insulator transition. 

It is instructive to compare the LAO/STO SLs to GTO/STO system~\cite{moetakef2012}. For the latter a similar critical thickness of $N=3$ was reported for the insulator-to-metal transition. In both cases the QW thickness reduction increases the effective carrier density and brings the system into a state where the on-site Coulomb repulsion becomes relevant, as pointed out by Ouellette \emph{et al.} \cite{Ouelette2013}.  

Concerning the orbital reconstruction most of previous studies have reported a dominant \dxy\ orbital polarization\cite{Salluzzo,pentcheva2006,pentcheva2008}. Here we find that this orbital polarization is stabilized only for a compressed $c$ lattice parameter, whereas the ground state comprises a \dxz, \dyz\ ordering at the \tith\ sites. Beyond the critical thickness, a crossover to a \dxy\ character in the interface layer and a mixed \dxz, \dyz\ character in the inner layers takes place. This is consistent with recent finding of a dominant contribution of \dxz, \dyz\ orbitals to the Fermi surface in LAO/STO, where doping levels beyond $\sim 6.9 \times10^{14}$ cm$^{-2}$ were achieved through oxygen vacancy doping \cite{petrovic2014}. The origin of orbital polarization and insulator-to-metal transition is further substantiated by the analysis of bond lengths (see. Fig.\ref{fig:Fig06}). The  \dxz, \dyz\ orbital polarization at the \tith\  sites is associated with significant elongation of the octahedra, whereas the transition from insulating-to-metallic state at the critical STO thickness is reflected in a reduction/increase in the apical to basal bond length ratio in the IF/IF-1 layer.

A key structural feature are the strong tilts of the TiO$_6$ octahedra, in particular at the interface,  in contrast to the undistorted bulk STO. In the related GTO/STO superlattices~\cite{moetakef2012} this behavior was ascribed to the influence of GTO, which has bulk B-O-B angles of 155$^\circ$. As LAO has a much larger B-O-B angle of 171.4$^\circ$, we can unambiguously conclude that the B-O-B angle reduction in the STO QW is not induced by the STO counterpart LAO, but correlates with the Ti $3d$ band occupation and the localization of charge. This is furthermore confirmed by the similar structural and electronic behavior of the  STO QW in the \ngo/\sto\  (bulk NGO has B-O-B angles of 153$^\circ$.) and the  \lao/\sto\ superlattices. In particular, the strongest changes in the in-plane B-O-B angles are observed at the interface at \tith\ sites, whereas the changes are much smaller and even quenched in the inner layers or at \tifo\ sites, as can be observed for $N=3$ and 4 in Figs. \ref{fig:Fig03}c and d.  

In summary, DFT+$U$ calculations, performed on (001)-oriented \lamstn\ and \ngmstn\ superlattices with $n$-type interfaces show a rich set of orbital reconstructions depending on the STO quantum well thickness and $c/a$ ratio. For $N=2$ the ground state is a charge ordered \tith, \tifo\ phase characterized by unanticipated \dxz,\dyz\ orbital polarization and elongated \tith-O$_6$ octahedra. $c$ lattice compression leads to a Dimer Mott insulator (\alao) or a \dxy\ orbital polarization (\asto), the latter case being the only one with compressed \tith-O$_6$ octahedra. With increasing STO quantum well thickness an insulator-to-metal transition takes place. The changes in the in-plane to out-of-plane bond length ratios signal the switching of orbital polarization and the critical thickness for the insulator-to-metal transition. A central finding is the pronounced enhancement of octahedral tilts in the STO quantum well that are not present in the bulk and can be unambiguously associated with the electrostatic doping of the polar $n$-type interface in these systems.   Together with the exotic electronic states found in (111)-oriented \lamstn\ SLs~\cite{doennig2013}, the results demonstrate how strain and the thickness of the STO quantum well can be used to engineer the orbital reconstruction and insulator-to-metal transitions in oxide superlattices.

\bigskip
{\Large {\bf Methods}}\bigskip

\label{sec:calc}
DFT calculations were performed on $n$-type \lamstn(001) and \ngmstn(001) superlattices with varying thickness $M$, $N$ of the constituents, using the all-electron full-potential linearized augmented-plane-wave (LAPW) method, as implemented in the WIEN2k code~\cite{wien2k}. For the exchange-correlation functional we used the generalized gradient approximation (GGA)~\cite{pbe96}. Static local electronic correlations were taken into account in the GGA+$U$ method~\cite{anisimov93} with $U=5$ eV, $J=0.7$ eV (Ti 3$d$), $U=8$ eV (La 4$f$, Nd 4$f$). For oxides containing \tith\ (3$d^1$) values between $U=3-8$ eV have been used in previous studies \cite{pentcheva2006,pentcheva2008,zhong2008}$^,$\cite{solovyev96,pavarini2004,streltsov2005}. The influence of strain was explored by setting the lateral lattice parameter to the one of LAO (\alao\ = 3.79\AA), NGO (\ango\ = 3.86\AA) or STO ($a_{\rm STO}^{\rm GGA}$ = 3.92\AA), corresponding to superlattices grown on the respective (001) substrate. Octahedral tilts and distortions were fully considered by relaxing atomic positions in P1 symmetry using a $\sqrt{2}a$ x $\sqrt{2}a$ lateral unit cell. Additionally the out-of-plane lattice parameter $c$ was optimized for all superlattices. The effective masses of the electrons were calculated by fitting the region around the Gamma point to a quadratic polynomial.

\bigskip
\emph{Note added to proofs}:

Such an emergence of strong octahedral tilts associated with a transition to a  Mott insulating phase  were recently predicted also in doped bulk SrTiO$_3$~\cite{Bjaalie2014}.

\begin{acknowledgments}
R.P. and D. D. acknowledge discussions with W. E. Pickett, S. Stemmer, Ariando, M. S. Golden and A. Janotti,  as well as financial support through the DFG SFB/TR80 (project C3/G3). Isosurface  plots of the electron density were generated by XCrySDen, written by A. Kokalj~\cite{xcrysden}.
\end{acknowledgments}

{\bf Author Contributions} 

D. D. performed the calculations under the guidance of R. P., D. D. and R. P. analyzed and interpreted the results and wrote the manuscript.  

{\bf Additional Information}

Competing financial interests: The authors declare no competing financial interests.

\end{document}